\documentclass[a4paper,11pt]{article}
\pdfoutput=1 % if your are submitting a pdflatex (i.e. if you have
\usepackage{jinstpub} % for details on the use of the package, please
\usepackage{subfigure}

\title{\boldmath Design of a TDC in SiGe technology for the front-end electronics of the RPC's  for high counting rate}

\author[a,b,1]{S. Bruno,}
\author[b,1]{R. Cardarelli,}
\author[a,b]{G. Aielli,}
\author[b]{D. Badoni,}
\author[a,b]{P. Camarri,}
\author[a,b]{A. Di Ciaccio,}
\author[a,b]{E. Alunno Camelia,}
\author[a,b]{A. Caltabiano,}
\author[a,b]{L. Massa,}
\author[a,b]{L. Pizzimento,}
\author[c]{L. Paolozzi,}
\author[a,b]{A. Rocchi}

\affiliation[1]{Authors}
\affiliation[a]{University of Rome TorVergata  }%Via della ricerca scientifica 1 , Roma, Italia}
\affiliation[b]{INFN Sezione Roma Tor Vergata\\Via della ricerca scientifica 1 , Roma, Italia}
\affiliation[c]{Universite de Geneve\\ Departement de Physique Nucleaire et Corpusculaire, Geneva, Switzerland  }

\emailAdd{salvatore.bruno@roma2.infn.it}

\abstract{ The new  generation of the RPCs is designed  to work with induced signals of few hundreds $\mu V$, hence the front-end electronics is an important and delicate part of the detector in order to get a detectable signal. The electronic chain described in this work is composed of an amplifier, a discriminator and  a TDC. The new front-end is realized in silicon-germanium (SiGe) BiCMOS technology, provided by IHP microelectronics. This technology implements BJT and CMOS transistors on the same chip. The benefit of this solution is to minimize the front-end power consumption ($2 \div 3$ $ \frac{mW}{ch}$) and  noise (500 $e^-$ r.m.s), while  improving the radiation hardness and  the response  speed of the electronics. In this work we will highlight the results from the first TDC prototypes. The TDC uses a local oscillator with an oscillation frequency ranging  between $0.6 \div 3.0 $  GHz and a time jitter of  15 ps. The data output from the TDC  are coded as binary numbers in order to lighten data processing to the acquisition system. The design of a serializer that sends the TDC data output to the acquisition system at 2 GHz is also described.

\begin{figure}[htbp!]
\centering
\includegraphics[scale=0.5]{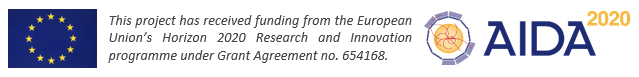}
\hspace{6mm}
\includegraphics[scale=0.2]{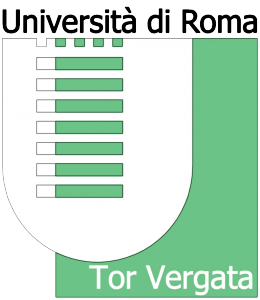}
\hspace{6mm}
\includegraphics[scale=0.5]{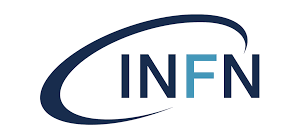}
\end{figure}

 }


\arxivnumber{1806.04082} % only if you have one

\proceeding{XIV RPC 2018 - Workshop on resistive plate chambers and related detectors\\
  19-23 February 2018\\
  Puerto Vallarta, Jalisco State, MEXICO}

\begin{document}
\maketitle
\flushbottom
\section{Motivations for the project and full-custom  TDC }
\label{sec:TDC}

\begin{figure}[htbp]
\centering
\includegraphics[scale=0.4]{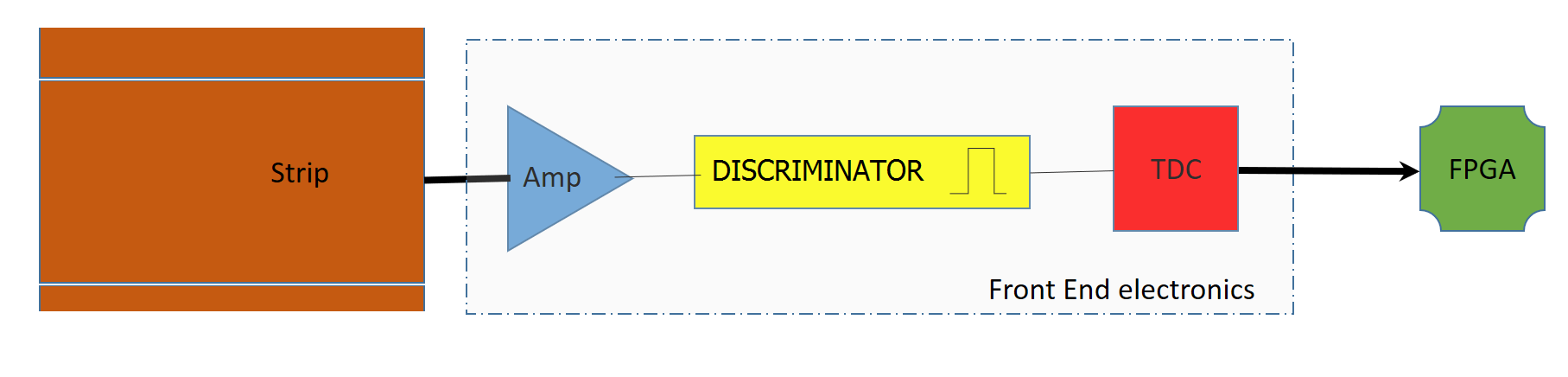}
\caption{Block Diagram of an RPC channel electronic chain. The front-end is realized within a single chip with the SiGe BiCMOS technology.}
\label{chain}
\end{figure}

%\section{Short motivations for the project}
%\label{sec:intro}
The RPC is a detector with an excellent time resolution.  To improve and optimize this parameter, an effective approach is to implement the complete front-end  chain in a single ASIC (figure \ref{chain}), using a SiGe BiCMOS technology \cite{Sigehbt,Sigebicmos}. Our design was implemented in SiGe 130nm technology by IHP microelectronics\footnote{ \emph{Innovations for High Performance microelectronics}; www.ihp-microelectronics.com}. The front-end uses a Time-to-Digital-Converter (TDC) for each electronic channel. \\
There are many advantages in the integration  of a full custum TDC inside the same chip with  the amplifier and discrimimator: lower cost, better time resolution compared to most commercial TDCs, intrinsic radiation-hardness of the technology, and low power consumption (less than $90\mu W/$channel). The BiCMOS SiGe technology has a better radiation tolerance than the classic Si BJT~\cite{SiGeATLAS,Radhard}:  the SiGe HBT technology tolerates gamma doses up to $500 kGy$ and a neutron dose of $10^{15} \frac{neq}{cm^2}$, much higher than the $10 kGy$ and $10^{13} \frac{neq}{cm^2}$ of the Si BJT technology \cite{Radhard}.\\

\begin{figure}[htbp]
\centering
\includegraphics[scale=0.4]{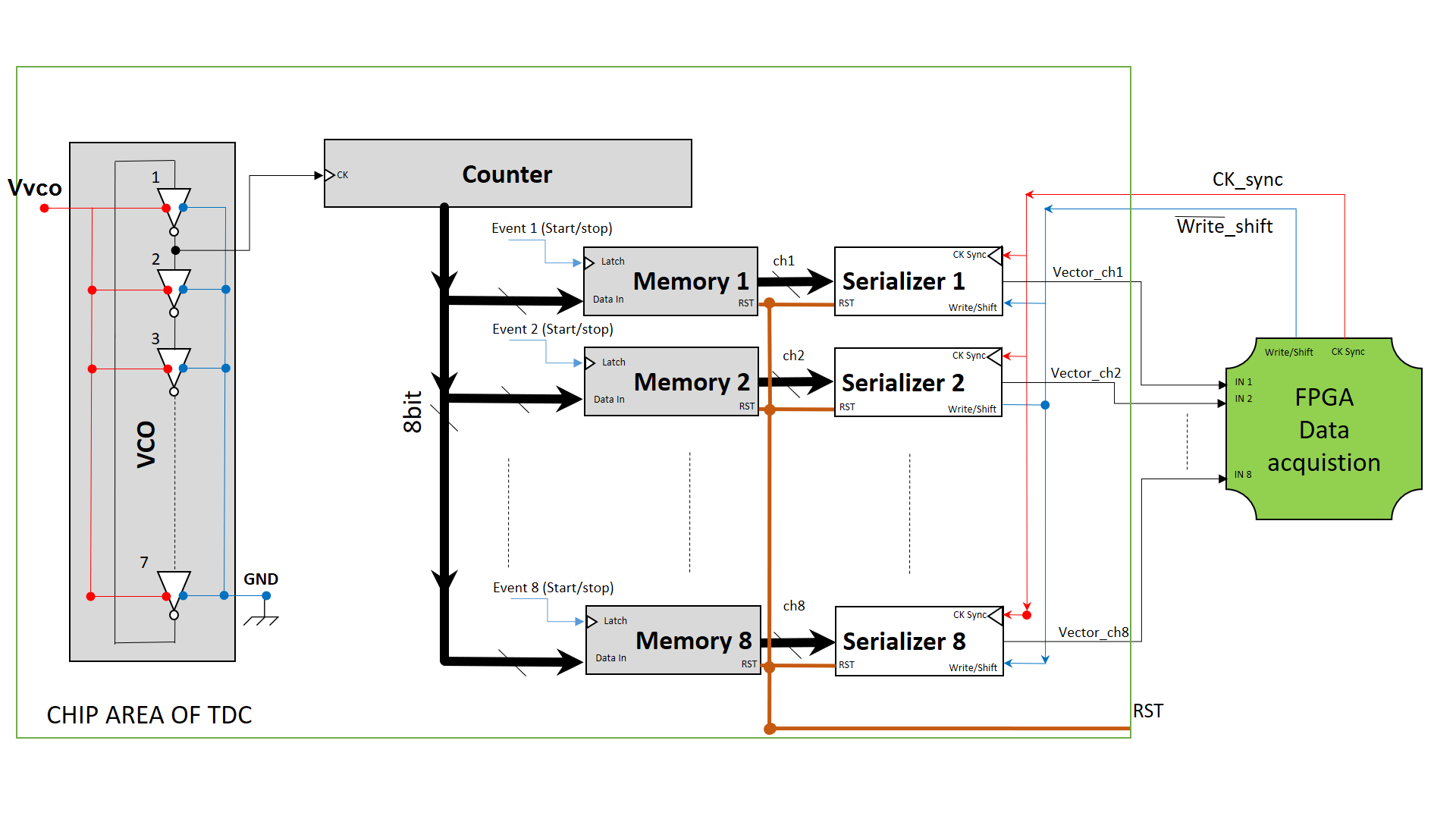}
\caption{ \label{blocco} Block diagram of the 8-channel TDC that will be installed inside the new  front-end electronics of the RPCs for high counting rate. The pins of the device are: two power supplies a power supply for all digital circuitry and a separated power for the VCO, for for reasons that will be explained in the next paragraph.; a safety \emph{reset} (RST); a \emph{synchronism clock}(\emph{CK\textunderscore{sync}}), which establishes the communication between the front-end and the FPGA; a $\overline{write}/shift$ used to read the status of the memories and send them to the FPGA. The eight \emph{Event} pins will be connected to the discriminator output. There are eight TDC pin output connected to the external FPGA, with are called in figure \emph{Vector \textunderscore{CH}}.}
\end{figure}

The schematic of the high-performance and very low-power TDC that we are developing, with a time resolution of $100  ps$ is shown in figure ~\ref{blocco}. It comprises an  internal Voltage Controlled Oscillator (VCO), a synchronous binary counter, eight memory blocks and a serializer that sends the data to the external FPGA.\\
The TDC logic is designed to work both on the rising and falling edge of the event.  The test results were obtained from the first TDC prototypes and are illustrated in the following paragraphs. 
In order to understand the results obtained from the prototype tests, it is necessary to explain the operating principles of the individual blocks.
 
\pagebreak 

\subsection{VCO}
\begin{figure}[htbp]
\centering % \begin{center}/\end{center} takes some additional vertical space
\includegraphics[width=.4\textwidth,trim=30 110 0 0,clip]{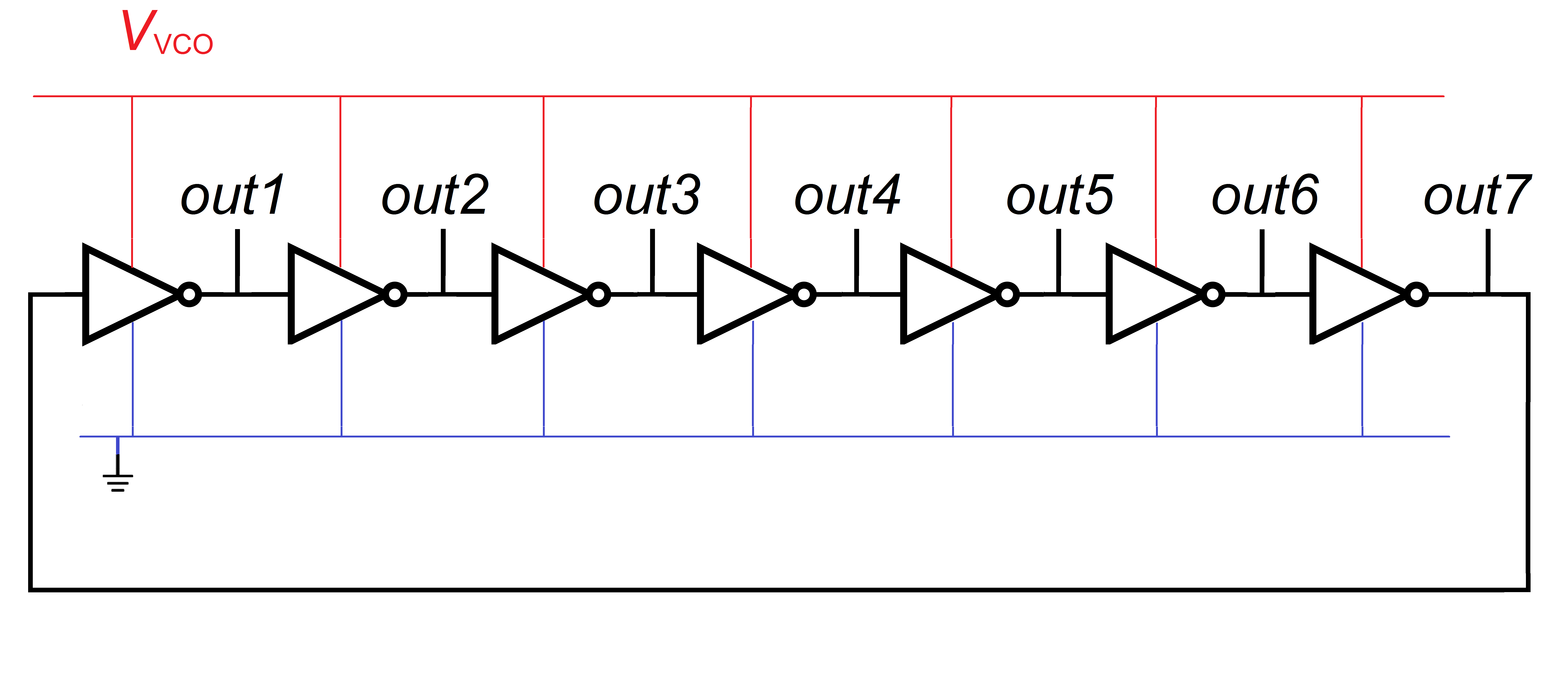}
\qquad
\includegraphics[width=.4\textwidth,trim=30 1 0 0,clip]{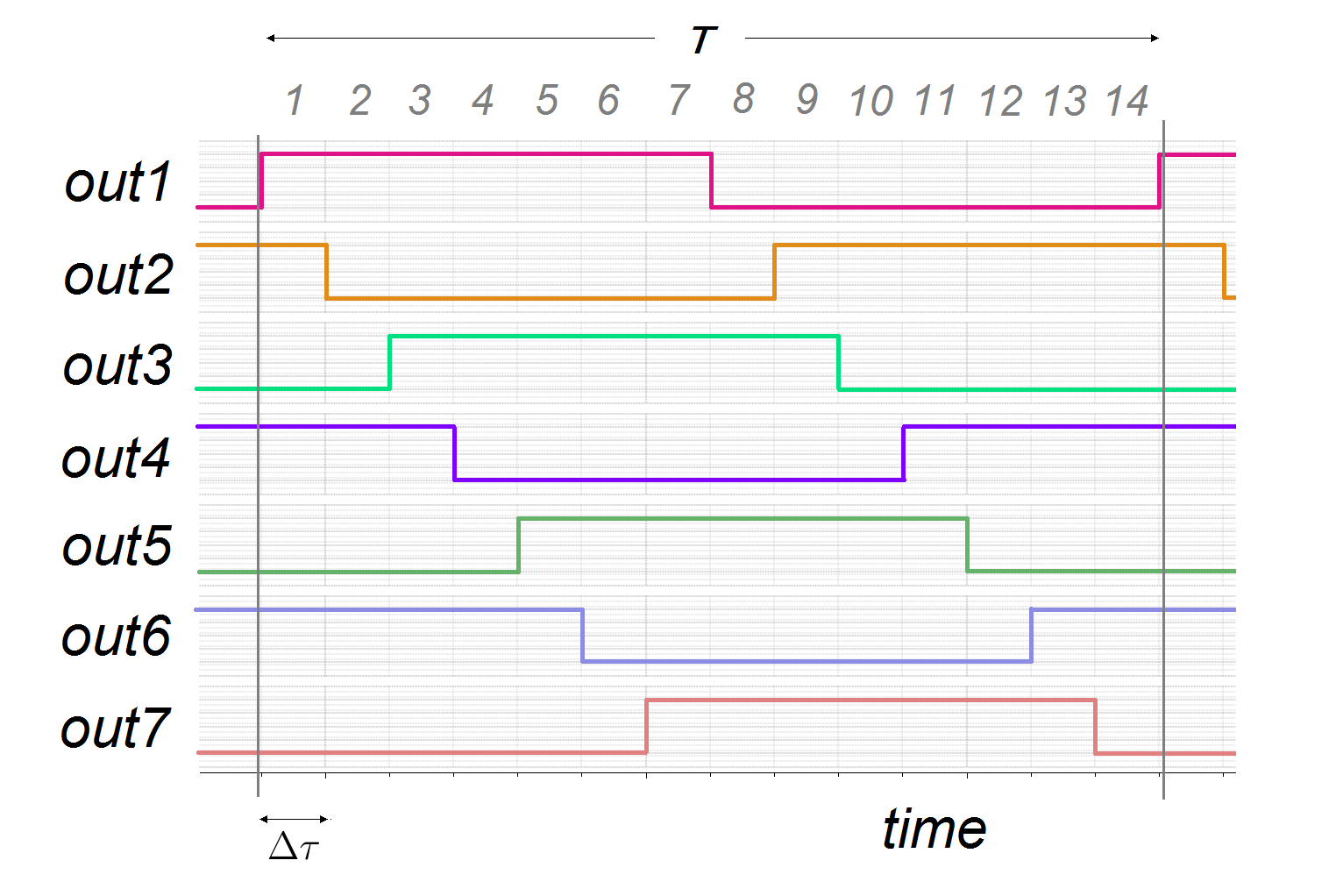}

\caption{\label{vco_fig} The figure on the left shows the schematic structure of the used VCO. The figure on the right shows the simulation result of the VCO output: each signal is delayed of $\Delta \tau$ with respect to the previous one.}
\end{figure}

The VCO is organized as a ring oscillator, (figure \ref{vco_fig} left) . For a given technology the transistors size and propagation time are defined thus, for a fixed number of inverters, the only free parameter that influences the frequency of oscillation is the supply voltage. In this condition, the ring oscillator can be used as a voltage controlled oscillator, using the supply voltage $ V_ {vco} $ as a frequency control parameter.
Assuming that the frequency is a linear function of the power supply voltage, we can write: 

\begin{equation}
F_{VCO}=f_0+K_{VCO}V_{vco}
\label{VCOFORMULA}
\end{equation}\\

where  $K_{VCO}$  is the gain coefficient of the VCO and $f_0$ is the minimum oscillation frequency. In  figure  \ref{vco_fig}, right, we can see the outputs of the seven inverter as a function of time. For a ring oscillator consisting in (2N+1) inverters, with a propagation delay of each of them of $ \Delta \tau $, the square wave has an oscillation frequency, $F_{VCO}$:
\begin{equation}
\centering
F_{VCO} = \frac{1}{2\ast (2N+1)\Delta\tau}
\label{freq_vco_eq}
\end{equation}

At the output of each inverter there is the same square wave, phase-shifted by $180^{\circ}$ and delayed of $ \Delta \tau $. This configuration of the VCO, realized with seven inverters, allows to distinguish 14 time intervals of duration $ \Delta \tau $ (figure \ref{vco_fig}, right side). \\

\begin{figure}[htbp]
\centering
\includegraphics[scale=0.4]{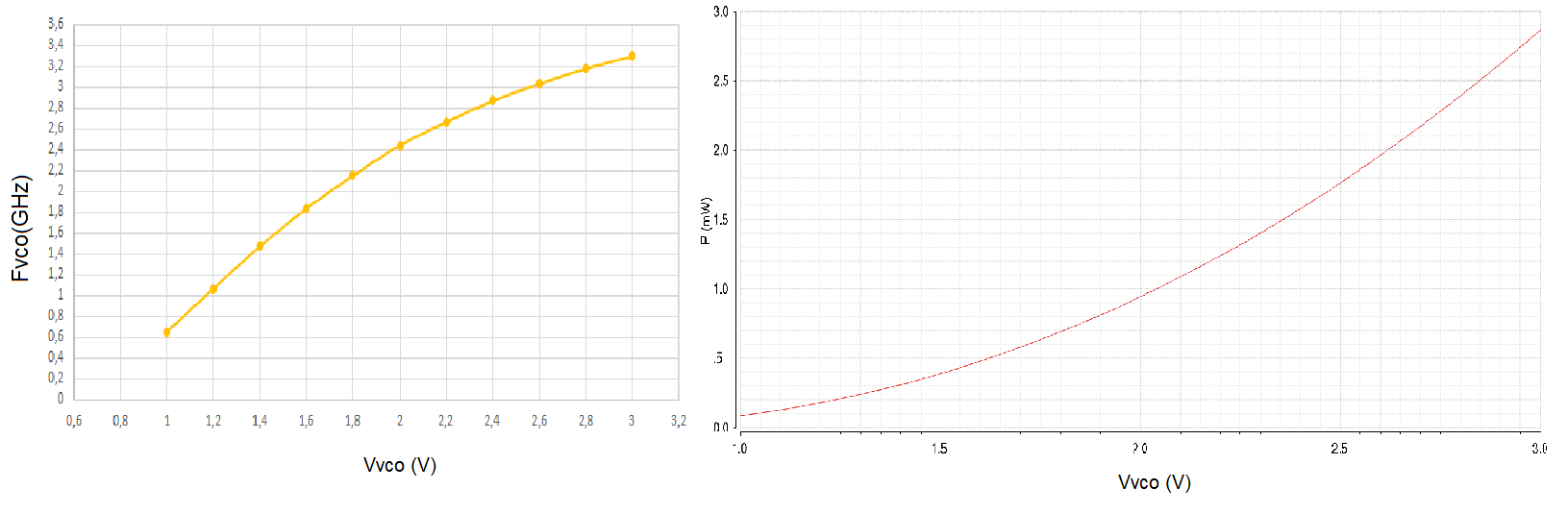}
\caption{\label{oscillazione_power_vco}  On the left,  mesurement of the  VCO oscillation frequency as a function of supply voltage. On the right,  power consuption of the VCO as a function of supply voltage..}
\end{figure}

The results form the measurements on the VCO (figure \ref{oscillazione_power_vco}, left), show how, varying the supply voltage, the oscillation frequency of the VCO can be set in range between 600MHz and 3.5GHz. 
The power consuption of the device in the same operation range is shown in figure \ref{oscillazione_power_vco}, right. 
	
 When working at the maximum oscillation frequency,  the  VCO  consumes less than 3mW. This is an excellent result as it allows to continue the development of the TDC to be inserted into the front end without adding cooling equipment.

\subsection{Binary Counter}
\label{parcounter}
\begin{figure}[htbp!]
\centering
\includegraphics[height=8.5cm, width=16cm]{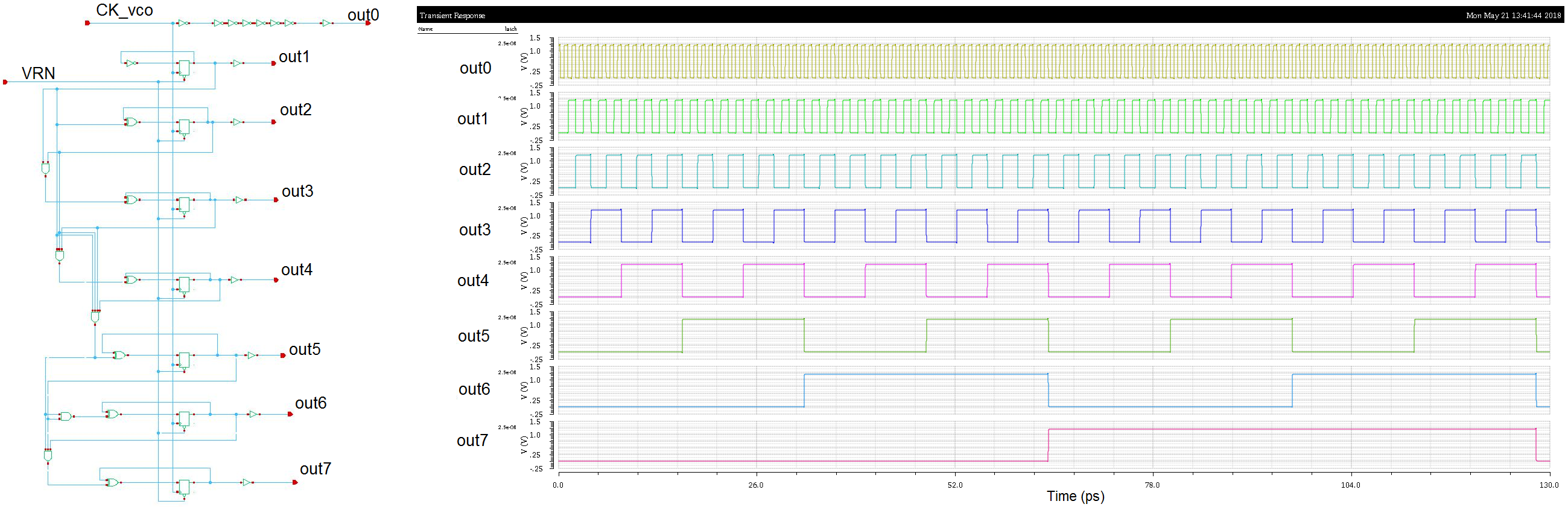}

\caption{\label{counter} On the left, electronic schematic of the 8-bit  synchronous counter. On the Right, simulation of the binary count where the output of the VCO are simulated with a periodic signal \emph{Ck \textunderscore {vco}} at a 1 GHz.}
\end{figure}

The counter block receives the output of the oscillator (figure \ref{blocco}). 
This counter is synchronous since the counting signal is applied simultaneously to all the flip-flops, with the same switching time(figure \ref{counter}).
This configuration guarantees a substantial independence from the switching time of the flip-flops. This block has the purpose to count the number of revolution of the VCO, the count is updated whenever the VCO makes a revolution. The counter has 8-bit of dynamic range and binary output encoding.

The counter can operate at a maximum frequency of 2 GHz, at a voltage of 1.5V. 

\subsection{Reading the memory block}
To perform the measurement we use a number of flip flops equal to the number of   outputs of the counter, (figure \ref{blocco}).  This block is used to read and save the counter outuput. The reading and saving is managed independently for each TDC channel by a signal called EVENT. For the prototypes under test the EVENT signal is fed externally from the chip. In the final chip, it will be 
the output of the disciminator of the corresponding channel.   Whenever there is an event the  flip-flop acquires the input data, making it available at the output.  The data is stored until the following event occurs. The memories will also have a maximum data acquisition rate. We have carried out a complete scan, measuring a maximum acquisition frequency of the memories of  2 GHz, when powered at 1.2V. 

\subsection{Serializer}
\label{parpiso}
\begin{figure}[htbp!]
\centering
\includegraphics[height=9cm, width=14cm]{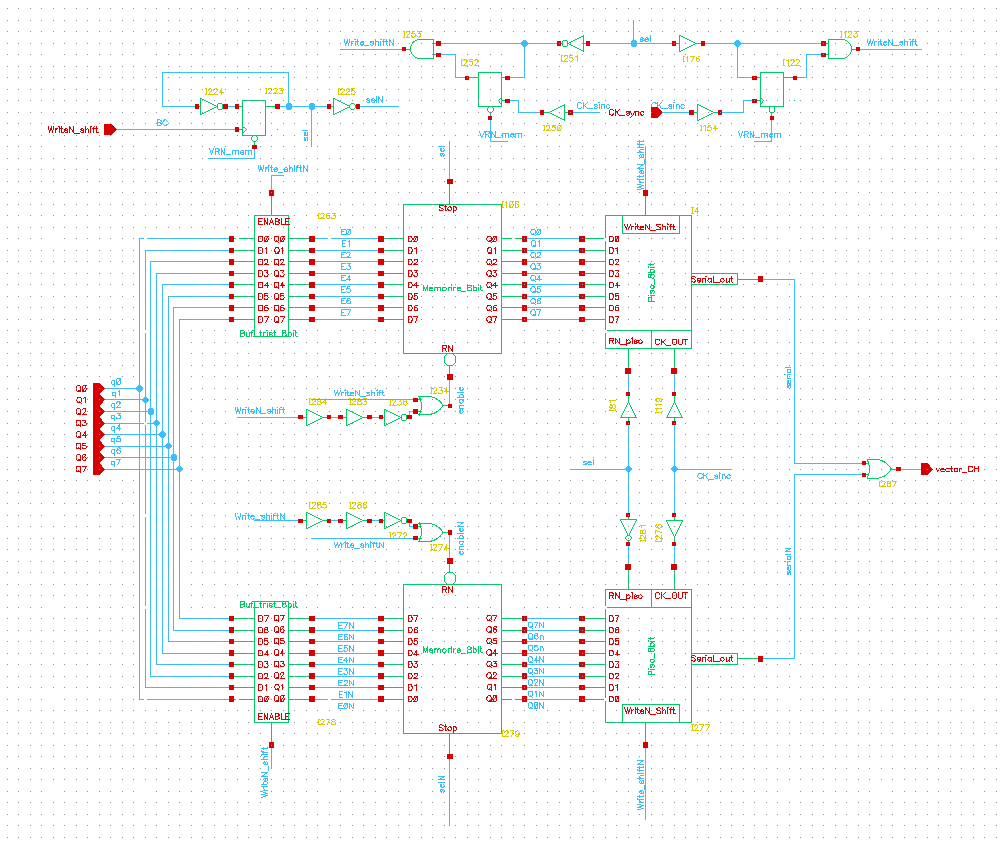}
\caption{\label{serial_upggrade} Serializer schematic. The design has been optimized to minimize the pile-up of the evets. }
\end{figure}

The last block of the schematic in figure  \ref{blocco} is the serializer. Its purpose is to convert the parallel input in a serial output, synchronize and send the data to the external FPGA. The input data of the  serializer comes from the memories described in the previous paragraph.
 It has a control signal $\overline{write}/shift$. When this control signal is to 0 logic, the data are read and saved. When it is  1 logic, the data are sent to external FPGA. The communication protocol between the front-end of the detectors and the FPGA is organized by a synchronism clock.
To reduce the pile-up of events, caused by the data transmission, we decided to divide the output of the memory block  in two different serializer channels (one for "even" events and one for "odd" events), figure \ref{serial_upggrade}. In this way the TDC has the capability to receive a large event rate because it will always have a serialier channel ready to save the data while the other one is sending data to the FPGA. The serializer permits to transfer an event rate up to the same clock frequency (CK\textunderscore {sync}) sent by the FPGA.

\section{Intrinsic Jitter and Power Consuptiond of a single TDC Channel }
\begin{figure}[htbp!]
\centering
\includegraphics[scale=0.5]{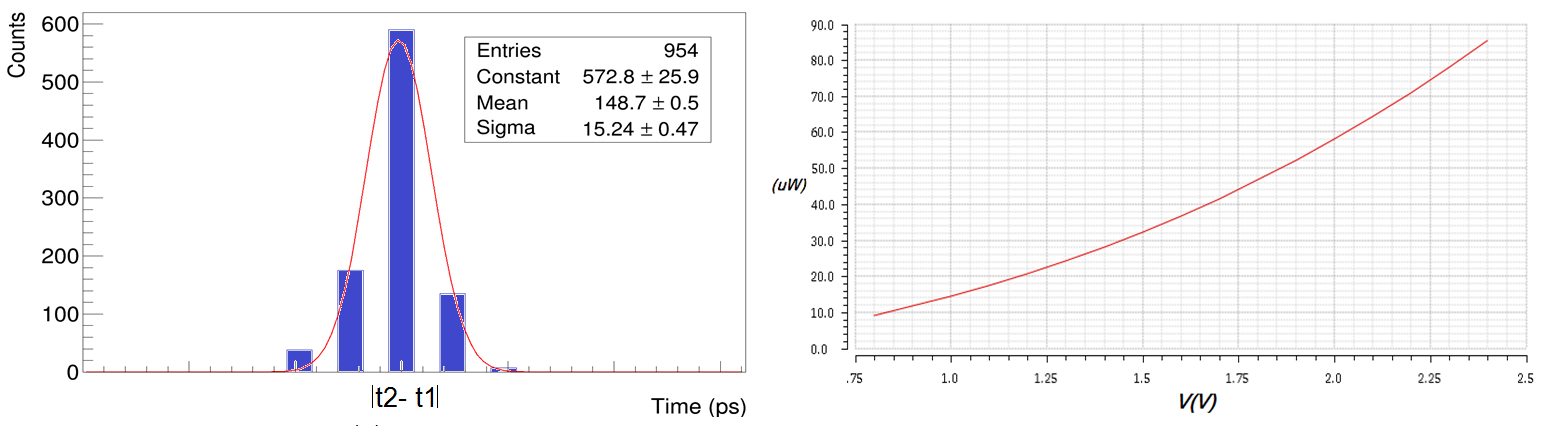}
\caption{\label{result} On the left, intrinsic jitter of the VCO obtained measuring the interval time between the rising edges of consecutive events produced at 100 MHz. On the right, power consuption of a single TDC channel, measured at an event rate of 100 MHz, constant $V_{vco}$ and varying the power supply of the rest of the circuit.. }
\end{figure}

As mentioned before, the technology used permits to reduce the power consumption of the front-end. Figure \ref{result} right, shows the power consumption of a single channel of the TDC, $10 \div 90 \mu W$, measured with an event rate of 100 MHz.
Another test that we have carried out consists in the measurement of the the intrinsic jitter of the VCO, (figure \ref{result}, left). This measurement gives the limit of the TDC precision. We measured the interval time between the rising edges of consecutive events, produced by a 100 MHz square wave generator with a precision of 5 ps. The intrinsic jitter of the VCO is $\sigma_{vco}=\frac{\sigma}{\sqrt{2}}=10.77ps$. This result  includes the internal jitter of the  pulse generator.

\section{Conclusions, observations and next steps}

The study of the implementation of a TDC in  SiGe technology at 130nm has given good results, which can be applied to  the  future RPC front-ends. We proved the feasibility of a full-custom TDC with the results from the prototypes. The intrinsic jitter of the VCO gives the possibility to realize a TDC with precision below 100ps.\\
The next step will be the creation of a new prototype to study the coupling between the discriminator and the TDC and optimize the critical parameters highlighted by the experimental data.

%\acknowledgments
%
%This is the most common positions for acknowledgments. A macro is
%available to maintain the same layout and spelling of the heading.
%\paragraph{Note added.} 


\end{document}